\documentclass{article}

\usepackage{arxiv}

\usepackage[utf8]{inputenc} 
\usepackage[T1]{fontenc}    
\usepackage{hyperref}       
\usepackage{url}            
\usepackage{booktabs}       
\usepackage{amsfonts}       
\usepackage{nicefrac}       
\usepackage{microtype}      
\usepackage{lipsum,amsmath,amssymb}		
\usepackage{graphicx}
\usepackage{doi}

\title{Extensibility governs the flow-induced alignment of polymers and rod-like colloids}

\author{\hspace{1mm}Vincenzo Calabrese\\
	Micro/Bio/Nanofluidics Unit\\ Okinawa Institute of Science\\ and Technology Graduate University \\Okinawa 904-0495, Japan\\ \texttt{vincenzo.calabrese@oist.jp} \\
	\And
	\hspace{1mm}Tatiana Porto Santos \\
	Micro/Bio/Nanofluidics Unit\\ Okinawa Institute of Science\\ and Technology Graduate University \\Okinawa 904-0495, Japan \\
 \And
	\hspace{1mm}Carlos G. Lopez \\
	RWTH Aachen University \\Institute of Physical Chemistry \\ Landoltweg 2, 52074 Aachen, Germany \\
 \And
	\hspace{1mm}Minne Paul Lettinga \\
	(\textit{a}) ICS-3, Institut für Weiche Materie \\Forschungszentrum Jülich, D-52425 Jülich, Germany\\ (\textit{b}) Laboratory for Soft Matter and Biophysics\\ KU Leuven, Celestijnenlaan 200D\\ B-3001 Leuven, Belgium
 \And
	\hspace{1mm}Simon J. Haward \\
	Micro/Bio/Nanofluidics Unit\\ Okinawa Institute of Science\\ and Technology Graduate University \\Okinawa 904-0495, Japan \\
  \And
	\hspace{1mm}Amy Q. Shen \\
	Micro/Bio/Nanofluidics Unit\\ Okinawa Institute of Science\\ and Technology Graduate University \\Okinawa 904-0495, Japan \\
}

\begin{document}
\maketitle

\begin{abstract}
Polymers and rod-like colloids (PaRC) adopt a favorable orientation under sufficiently strong flows. However, how the flow kinematics affect the alignment of such nanostructures according to their extensibility remains unclear. By analysing the shear- and extension-induced alignment of chemically and structurally different PaRC, we show that extensibility is a key determinant of the structural response to the imposed kinematics. We propose a unified description of the effectiveness of extensional flow, compared to shearing flow, at aligning PaRC of different extensibility.
\end{abstract}

Polymers and rod-like colloids (PaRC) are ubiquitous in biological fluids (e.g., mucus, saliva), food and industrial formulations (e.g., gels, paints), imparting specific properties and functionalities. When solubilized or dispersed in a solvent, (PaRC) adopt an equilibrium conformation (e.g., rod-like, worm-like, coil-like) that depends on a multitude of factors including surface chemistry, contour length, and backbone rigidity, affecting the PaRC extensibility and flexibility~\cite{rubinstein2003polymer}.
The PaRC extensibility ($L_e$) describes the ratio between the contour length of the fully extended structure ($l_c$) and the respective size at equilibrium. The PaRC flexibility ($N_p$) represents the number of persistence length ($l_p$) segments composing the PaRC contour length~\cite{rubinstein2003polymer,doi1988Book}.
For rigid colloidal rods $L_e\rightarrow1$ and $N_p\ll1$, whilst for flexible polymers that adopt a coil-like conformation at equilibrium, $L_e\gg1$ and $N_p \gg 1$.  Under sufficiently strong imposed flows, PaRC are driven out of equilibrium and towards a state of alignment induced by velocity gradients in the flow field. Flow-induced alignment of PaRC is central to processes including fiber spinning~\cite{rosen2021elucidating}, bacteriophage replication~\cite{chen2018two,grayson2007real}, and amyloid fibrillogenesis related to neurodegenerative diseases~\cite{hill2006shear,akkermans2008formation}. The link between the PaRC conformation and the bulk fluid properties at the equilibrium, such as the zero-shear viscosity and the macromolecular time scales of diffusion and relaxation, are well understood and described by generalized scaling theories~\cite{rubinstein2003polymer,Colby2010,doi1988Book}. Nonetheless, how the dynamics of PaRC under flow depend on their equilibrium conformation remains elusive. 

It is known that rigid colloidal rods tend to adopt a favourable orientation under characteristic deformation rates $|E|$ able to overcome the rotational Brownian diffusion~\cite{doi1988Book,calabrese2023naturally,Lang2019}. This condition is captured by the P\'eclet number $Pe=6|E|\tau \gtrsim1$, with $\tau$ the longest relaxation time of the rigid rod~\cite{doi1988Book,Doi1978P1}. For sufficiently strong shear flows ($Pe\gtrsim1$), rigid colloidal rods orient but also undergo occasional tumbling due to the vorticity component of the strain rate tensor~\cite{Dhont2007,zottl2019dynamics}. On the contrary, pure extensional flows are vorticity-free and generally regarded as more effective than shear flows at inducing the alignment of rigid colloidal rods~\cite{Calabrese2021,Calabrese2022b,Corona2022a,Chun2023a,calabrese2023naturally}. For flexible polymers in solution, favourable chain orientation is expected to occur at Weissenberg number $Wi=|E| \tau \gtrsim 0.5$, with $\tau$ the longest relaxation time of the polymer~\cite{DeGennes1974a,Keller1985,Schroeder2018,larson1989coil}. 
Extensional flows are considered pivotal at inducing a preferential orientation of flexible polymers. Direct single molecule imaging of $\lambda\text{-DNA}$ ($L_e\sim 13$) has shown that extensional flows provide a greater extent of chain extension than shear flows~\cite{smith1999single,Hur2002}. For solutions of even more highly flexible synthetic polymers (e.g., poly(ethylene oxide) and atactic-polystyrene with $L_e> 50$) microfluidic studies reporting flow-induced birefringence (FIB) show that macromolecular alignment is dominant in localized regions originating from hyperbolic points in the flow field where purely extensional kinematics are approximated~\cite{Haward2012a,Haward2016,Keller1985,fuller1995optical}.

In this letter we investigate the steady shear- and extension-induced alignment of PaRC with varying extensibility, including rod-like colloids and significantly more flexible polyelectrolytes in solution. Employing quantitative FIB imaging in a specific microfluidic device, we show that different deformation rates are required for the onset of PaRC alignment in shear and extension, and that the difference is correlated with the PaRC extensibility. In the limit of rigid rod-like nanostructures we validate our results with the revised Doi-Edwards theory proposed by Lang et al.\cite{Lang2019} Our study aims towards unifying the relationship between shear- and extension-induced alignment of PaRC with various extensibility.
\begin{figure}[t!]
    \centering
    \includegraphics[width=8.6cm]{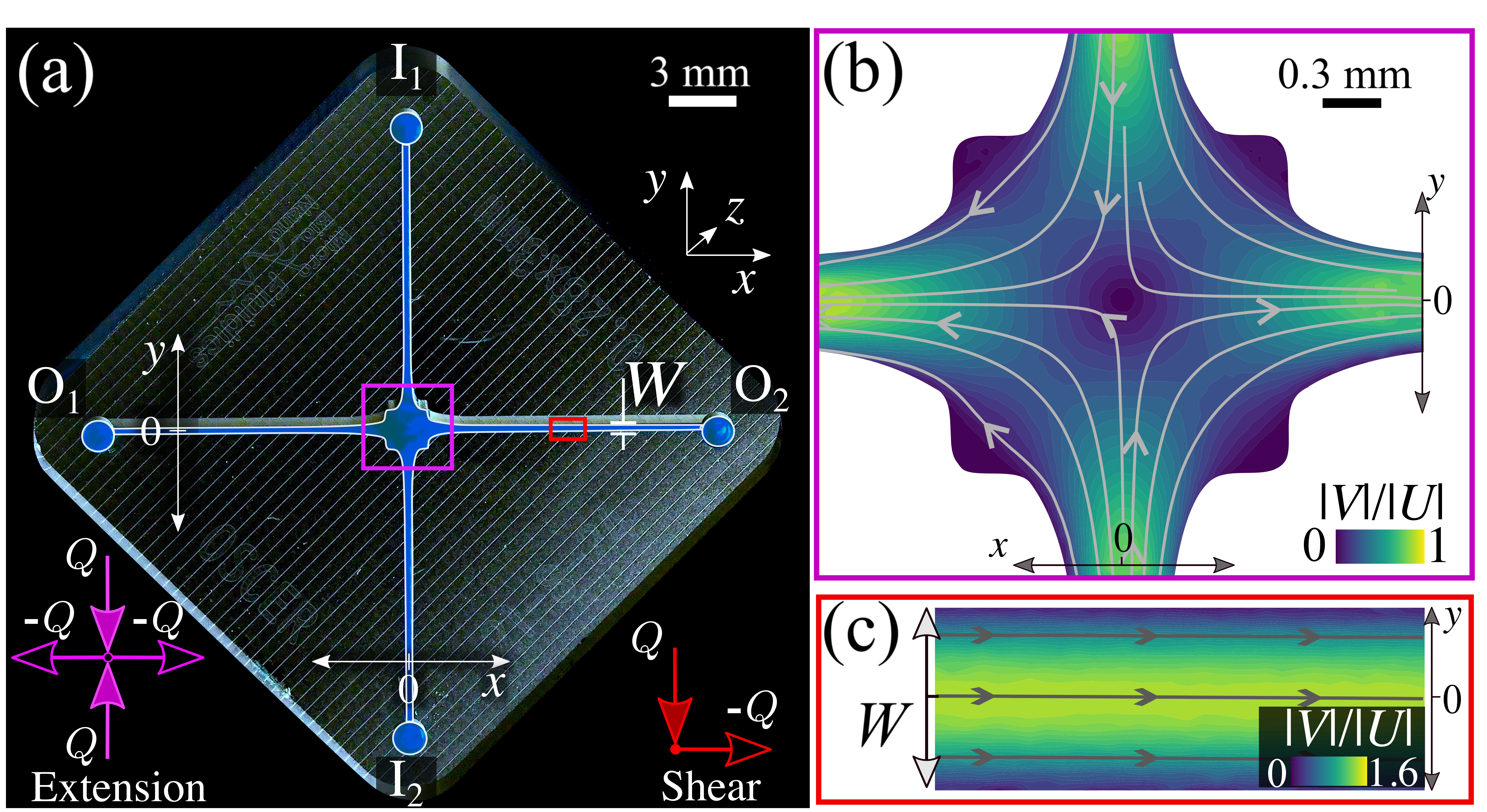}
    \caption{(a) A photograph of the OSCER device with the extension- and shear-dominated ROI marked by the magenta and red box, respectively. Flow fields for a Newtonian fluid at $Re<1$ in the extension- and shear-dominated ROI in (b) and (c), respectively. The measured velocity magnitude $|V|$ is scaled by the average flow velocity $|U|$.}
    \label{fig:sketch}
\end{figure}

PaRC alignment is evaluated via FIB imaging performed on a CRYSTA PI-1P camera (Photron Ltd, Japan) fitted with a 4$\times$ Nikon objective lens. Microparticle image velocimetry (\textmu PIV) (TSI Inc., MN, 4$\times$ and 10$\times$ Nikon Objectives) is used to characterize the nature of the imposed flow kinematics in the microfluidic device~\cite{supp}. The microfluidic device (Fig.~\ref{fig:sketch}(a)), fabricated in fused silica glass~\cite{Burshtein2019}, is a numerically-optimized version of the cross-slot device, referred to as the optimized shape cross-slot extensional rheometer (OSCER)~\cite{Haward2012a,Haward2016}. The device has an aspect ratio $H/W=10$, where $H=3~\text{mm}$ is the height, defining the optical path along the $z-$axis, and $W=0.3~\text{mm}$ is the width (in the $x\text{-}y$ plane), producing a good approximation to a two dimensional flow. FIB and \textmu PIV are evaluated at the $x$-$y$ plane in two distinct regions of interest (ROI) of the microfluidic device generating extension- or shear-dominated flows (marked by the coloured boxes in Fig.~\ref{fig:sketch}(a)). In the extension-dominated ROI (magenta box, Fig.~\ref{fig:sketch}(a)), FIB and \textmu PIV are evaluated around the stagnation point (i.e., $x=y=0~\text{mm}$) where a constant and shear-free planar extension is generated when opposing inlets ($\text I_1$, $\text I_2$) and outlets ($\text O_1$, $\text O_2$) operate at equal and opposite volumetric flow rates ($Q~\text{m}^3 /\text{s}$) (representative flow field in Fig.~\ref{fig:sketch}(b)). A shear-dominated ROI (red box, Fig.~\ref{fig:sketch}(a)) located in one of the OSCER branches is evaluated while one inlet ($\text I_1$) and outlet ($\text O_2$) operate at an equal and opposite flow rate, generating a shear-dominated flow without a stagnation point (representative flow field in Fig.~\ref{fig:sketch}(c)). Flow is controlled by syringe pumps (Nemesys, Cetoni) to impose an average flow velocity $|U|=Q/(HW)$ and, typically, a Reynolds number $Re =\rho |U| W/\eta\lesssim 1$ , with $\rho$ the fluid density and $\eta$ the shear viscosity at a nominal shear rate $Q/(W^2 H$)~\cite{supp}.

The PaRC investigated in this work include rod-like colloids that are cellulose nanocrystals (CNC), oxidized cellulose nanofibrils (OCNF), protein nanofibrils (PNF), and two types of filamentous viruses (Pf1 and fd)~\cite{supp}. We use carboxymethyl cellulose sodium salt (CMC) and hyaluronic acid sodium salt (HA) with distinct molecular weight (Mw), and double stranded calf-thymus DNA sodium salt (ct-DNA) with $\sim 28~\text{kbp}$ as polyelectrolytes in solution~\cite{supp}. CMC with Mw of 0.4, 0.6 and 0.8~MDa, are referred to as CMC04, CMC06 and CMC08, respectively, and HA with Mw of 0.9, 1.6, 2.6 and 4.8~MDa, are referred to as HA09, HA16, HA26 and HA48, respectively~\cite{Haward2014a}. For the rod-like colloids, the characteristic contour length $\langle l_c \rangle $ and the persistence length $l_p$ are obtained from atomic force microscopy (AFM)~\cite{Usov2015a,supp}. For CMC and HA, $l_p$ is obtained from literature~\cite{lopez2018electrostatic,norisuye2000conformation,salamon2013probing}, and $\langle l_c \rangle = N a$, where $N$ is the is the weight averaged degree of polymerisation, and $a$ is the monomer length.\cite{supp} For ct-DNA, $\langle l_c \rangle$ is obtained by multiplying the number of base pairs with their average separation distance ($0.34~\text{nm}$) and  $l_p=50~\text{nm}$~\cite{Schroeder2018,reisner2012dna}. The concentrations of the PaRC are chosen to be as dilute as possible yet sufficient to provide a detectable birefringence signal. Where required, the relaxation time of PaRC, $\tau$, is increased by adding glycerol or sucrose to the acqueous solvent, shifting the onset of PaRC alignment within the experimental window of accessible deformation rates. The extensibility of the PaRC is computed as $L_e=\langle l_c \rangle / \sqrt{\langle R^2 \rangle}$ where $\langle R^2 \rangle = 2 l_p \langle l_c \rangle - 2 l_p^2 (1 - \mathrm{exp} (-\langle l_c \rangle /l_p))$ is the mean-squared end-to-end length of PaRC at equilibrium using a worm-like chain model~\cite{rubinstein2003polymer}. The critical length scale $\ell$ below which PaRC follow a rod-like behaviour (i.e., $\sqrt{\langle R^2 \rangle}\propto \ell$) denotes the persistence length $l_p$. The PaRC flexibility $N_p=\langle l_c \rangle /l_p$, is related to $L_e$ via:  
\begin{equation}
L_e=\sqrt{\frac{N_p^2}{2(N_p+\text{exp}(-N_p)-1)} },
\label{eqn.LeNp}
\end{equation}
which approaches the scaling $L_e\propto N_p^{0.5}$ for large $N_p$.
\begin{figure*}
    \centering
    \includegraphics[width=16cm]{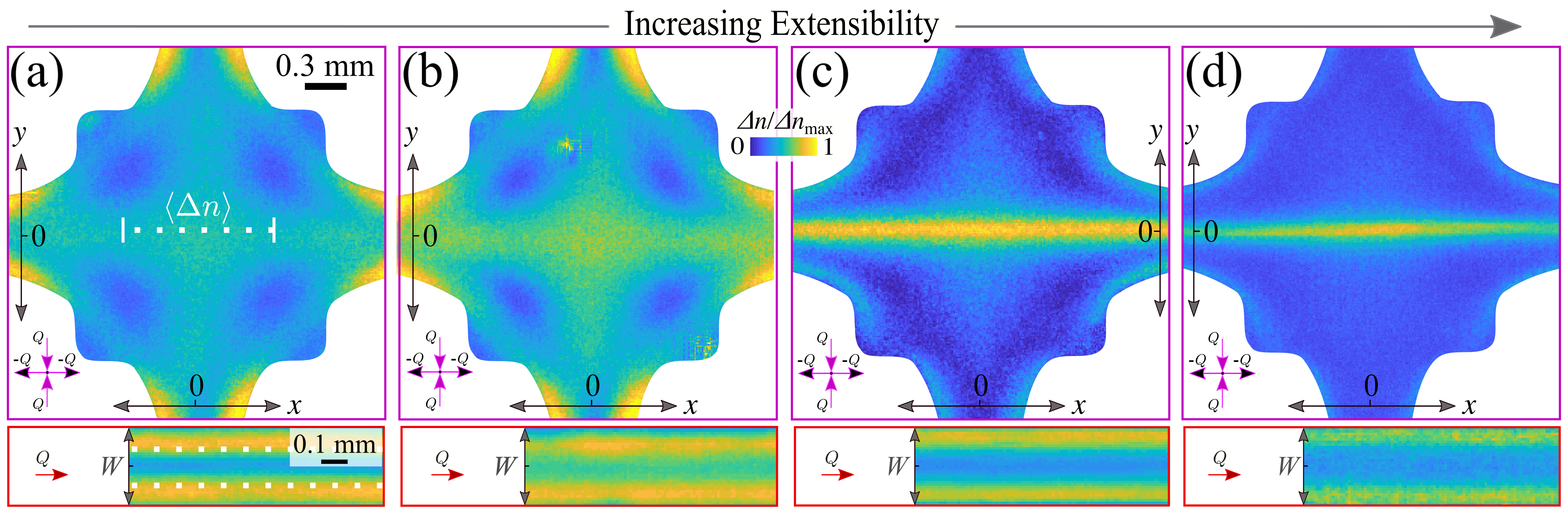}
    \caption{Steady state birefringence patterns for PaRC with different $L_e$ at an equal average flow velocity ($|U|$) in the extension- and shear-dominated ROI (top and bottom panels, respectively). For each system and ROI, the birefringence ($\Delta n$) is normalized by the maximum birefringence ($\Delta n_{\text{max}}$). (a)  fd dispersion (0.2 mg/mL) at $|U|=3.6~\text{mm/s}$. (b) Pf1 dispersion (0.1 mg/mL) at $|U|=1.9~\text{mm/s}$. (c) CMC08 solution (0.1 mg/mL) at $|U|=95.7~\text{mm/s}$. (d) HA48 solution (1 mg/mL) at $|U|=1.9~\text{mm/s}$.}
        \label{fig:BirefringencePattern}
\end{figure*}

FIB patterns in the extension- and shear-dominated ROI acquired at an equal average flow velocity $|U|$ are shown in Fig.~\ref{fig:BirefringencePattern} for representative PaRC with distinct $L_e$. For each system and ROI investigated, the birefringence ($\Delta n$), probing the extent of fluid anisotropy due to the PaRC alignment, is normalized by the maximum birefringence ($\Delta n_{\text{max}}$). For fd and Pf1 a diamond-like pattern of intense birefringence is shown in the extension-dominated ROI as previously shown for colloidal rods with $L_e\sim 1$~\cite{Calabrese2021,Calabrese2022b}. For the comparatively more extensible CMC08 ($L_e\sim 7$), the diamond-like birefringence pattern fades and an intense birefringence strand develops along the extensional axis (i.e., $x$). For the most extensible PaRC investigated, HA48 ($L_e\sim 16$), the diamond-like birefringence pattern disappears and only the intense birefringence strand becomes visible. Fluid elements accumulate strain exponentially as they enter the extensional ROI and approach the extension axis~\cite{Haward2016}. Thus, more extensible PaRC need to pass close to the stagnation point to accumulate enough strain to stretch, leading to a localized birefringence strand. Contrarily, more rigid rod-like PaRC require a relatively small accumulated strain to align, leading to a significant alignment even away from the stagnation point, resulting in a diamond-like birefringence pattern. In the shear-dominated ROI, the birefringence increases from the centerline to the channel walls, in qualitative agreement with the trend of shear rate $|\dot\gamma|$ retrieved from \textmu PIV~\cite{supp}. 
To quantify the difference between shear- and extension-induced alignment of PaRC with distinct extensibility, we spatially average the birefringence ($\langle \Delta n \rangle$) in specific locations of the shear- and extension-dominated ROI where the shear and extension deformation rates are constant. In the extension-dominated ROI, the birefringence is averaged along 1~mm of the extension axis (i.e., at $y=0$ for $-0.5<x<0.5$, dotted line in Fig.~\ref{fig:BirefringencePattern}(a), top panel) and plotted as a function of the extension rate $|\dot\varepsilon|$ along the same line (determined from the flow field measured by \textmu PIV). In the shear-dominated ROI, the birefringence is averaged along the $x$ direction at $y=\pm W/4$ (marked by the dotted line in Fig.~\ref{fig:BirefringencePattern}(a), bottom panel) and plotted as a function of the respective shear rate at the same location ($|\dot\gamma|$)~\cite{supp}. We note that the same flow strength for shear and planar extension is given by the magnitude of the strain rate tensor as $|\dot\gamma|=2|\dot\varepsilon|$. 
\begin{figure*}[!]
\centering        \includegraphics[width=16.5cm]{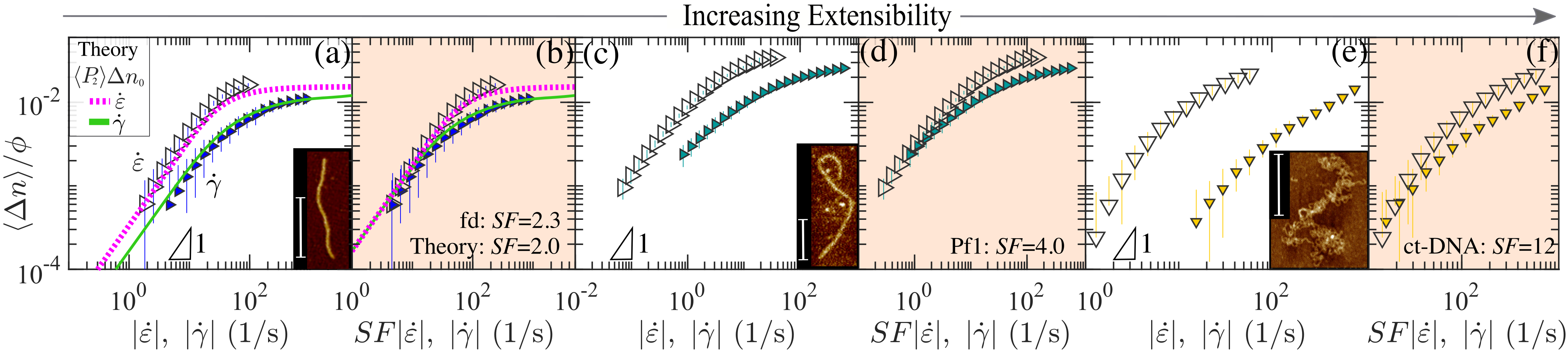}
    \caption{Spatially averaged birefringence ($\langle \Delta n \rangle$) normalized by the mass fraction of the PaRC ($\phi$) as a function of the extension rate ($|\dot\varepsilon|$) (empty symbols) and shear rate ($|\dot\gamma|$) (filled symbols). (a, b) fd dispersion at 0.2~mg/mL. (c, d) Pf1 dispersion at 0.5 mg/mL. (e, f) ct-DNA solution at 0.19~mg/mL. In (b), (d), and (f) the $\langle \Delta n \rangle / \phi$ as a function of $|\dot\varepsilon|$ is re-scaled as $SF|\dot\varepsilon|$. In (a) we provide direct comparison with the order parameter, $\langle P_2 \rangle$, vs. $|\dot\gamma|$ and $|\dot\varepsilon|$ from the rigid rod theory~\cite{Lang2019} where $\langle P_2 \rangle$ is scaled as $\langle P_2 \rangle \Delta n_0$. The insets in (a), (c) and (e) are representative AFM images of fd, Pf1, and ct-DNA with a scale bar of 500~nm. }
    \label{fig:Birefringence}
  \end{figure*}

In Fig.~\ref{fig:Birefringence} we plot the normalized birefringence, $\langle \Delta n \rangle/\phi$, with $\phi$ the mass fraction of the PaRC, as a function of $|\dot\gamma|$ (filled symbols) and $|\dot\varepsilon|$ (empty symbols) for three representative PaRC with distinct $L_e$. For the relatively rigid fd ($L_e\sim 1$), the onset of birefringence occurs at lower values of $|\dot\varepsilon|$ than $|\dot\gamma|$, indicating that fd alignment is induced more readily by extension than by shear (Fig.~\ref{fig:Birefringence}(a)). 
We introduce a non-dimensional scaling factor ($SF$) to the extension rate as $SF|\dot\varepsilon|$ to match the onset of birefringence in extension with that in shear. Practically, $SF$ provides an estimate of the ratio between the critical shear rate ($|\dot\gamma^*|$) and critical extension rate ($|\dot\varepsilon^*|$) at the onset of PaRC alignment ($SF=|\dot\gamma^*|/|\dot\varepsilon^*|$). For fd, $SF=2.3$ captures the difference between extension- and shear-induced alignment at low deformation rates (Fig.~\ref{fig:Birefringence}(b)).
 The scaling procedure highlights the difference at high deformation rates where the birefringence in extension reaches a greater plateau value than observed in shear. Using the known dimensions of the fd virus, we compare our experimental results with the revised Doi-Edwards theory for ideal, rigid and monodisperse rods proposed by Lang et al.~\cite{Lang2019,supp} The theoretical projected order parameter $\langle P_2\rangle$ is compared with the experimentally measured birefringence as $\langle P_2 \rangle \Delta n_0= \langle \Delta n \rangle / \phi$, where $\Delta n_0$ is the birefringence of perfectly aligned PaRC at $\phi=1$~\cite{Purdy2003,Uetani2019,RosenPRE2020}. To provide a close comparison between theory and the experimental data, in Fig.~\ref{fig:Birefringence}(a, b) the theoretical curves are scaled by $\Delta n_0=0.015$, a value comparable to a previously reported value for fd~\cite{Purdy2003}. The theory predicts well the birefringence increase with a slope of 1 in both shear and in extension and also predicts the saturation of birefringence at high deformation rates (Fig.~\ref{fig:Birefringence}(b)). The theory confirms that (i) a lower extension rate is required to induce fd alignment compared to the shear rate, and that (ii) at high deformation rates the greatest extent of alignment occurs in extensional flow. Both (i) and (ii) are consistent with the absence of tumbling events in extension, enhancing the overall extent of alignment. For the theoretical curves, $SF=2$ (i.e., comparing equal flow strength) enables superimposition of the birefringence curves at low deformation rates in a similar fashion as for the fd dispersion (Fig.~\ref{fig:Birefringence}(b)). Thus, on the base of theory  $SF=2$ is expected in the limiting case of rigid rod-like nanostructures ($L_e=1$). The slight difference between $SF$ from the experiment (e.g., $SF=2.3$ for fd) and theory is potentially associated with the fact that fd is not perfectly rigid as assumed by the theory (see inset AFM image in Fig.~\ref{fig:Birefringence}(a)).  

\begin{figure}[t!]\centering
    \includegraphics[width=8.6cm]{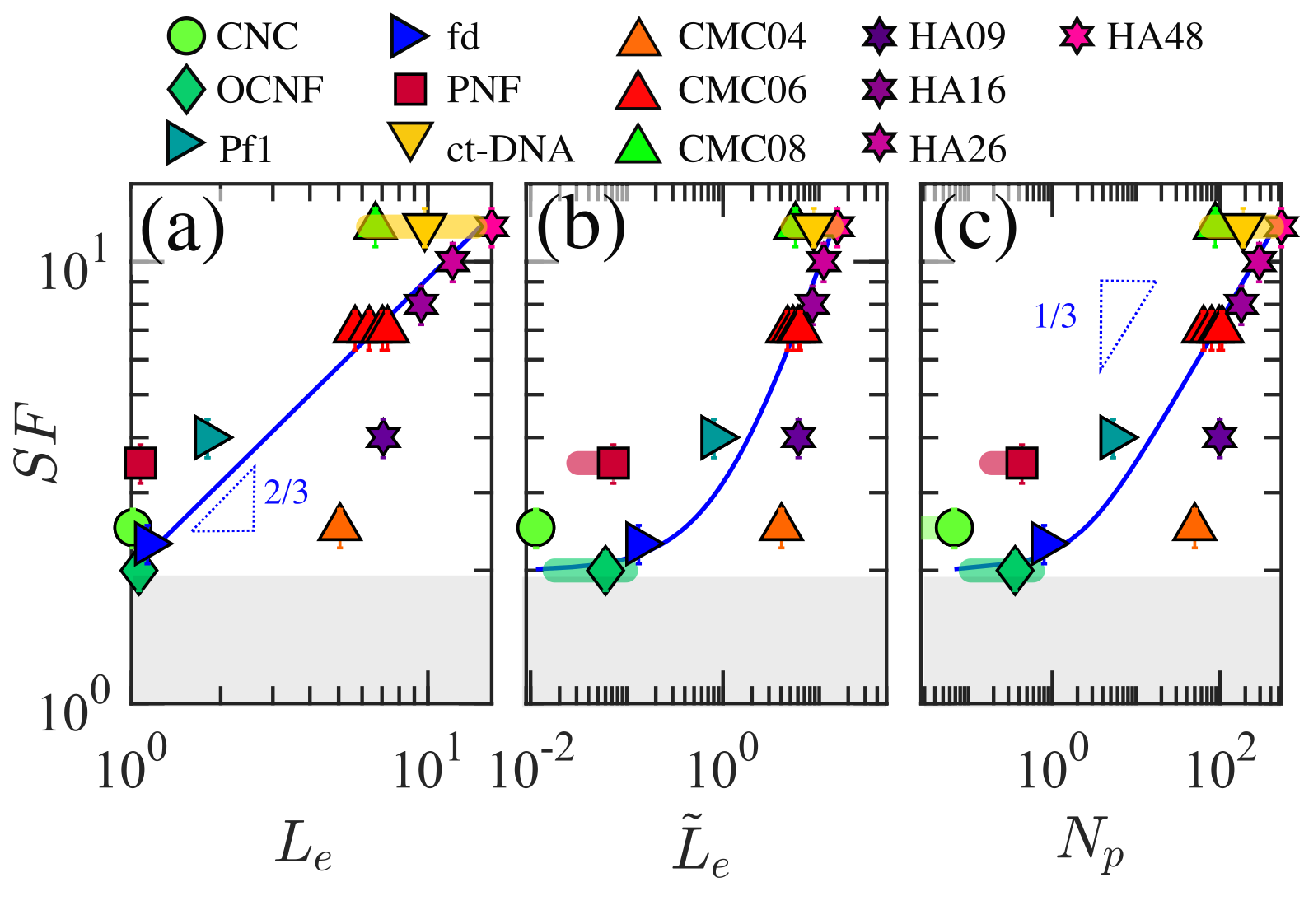}
\caption{$SF$ as a function of $L_e$, $\tilde L_e$ and $N_p$, in (a), (b) and (c), respectively. The solid line in (a) is describes $SF=2L_e^\beta$. In (b) the solid line is the plot of $SF=2(\tilde L_e + 1)^\beta$ and in (c) is the plot of $SF=2\bigg(\sqrt{\frac{N_p^2}{2(N_p+\text{exp}(-N_p)-1)} }\bigg)^\beta$. The gray area sets the threshold $SF=2$ for $L_e= 1$ based on rigid rod theory. For CNC, OCNF, PNF and ct-DNA the shadowed segments represent the boundaries of $Le$, $\tilde Le$ and $N_p$ based on the measured contour length distribution~\cite{supp}. }
    \label{fig:Scaling}
\end{figure}

With progressively increasing extensibility, the difference between $\lvert \dot\varepsilon^*\rvert$ and $\lvert \dot\gamma^*\rvert$ becomes more pronounced, as shown for Pf1 ($L_e\approx2$, Fig.~\ref{fig:Birefringence}(c, d)) and ct-DNA ($L_e\approx 10$, Fig.~\ref{fig:Birefringence}(e, f)). In these cases, scaling factors of $SF=4$ and $SF=12$ are required, respectively, to match the onset of birefringence. These results suggest that with the increasing PaRC extensibility, extensional deformations become progressively more effective at inducing the onset of PaRC alignment than shear deformations. Since the effectiveness of the extensional deformations relative to shear deformations at inducing PaRC alignment is captured by $SF$, we compare the $SF$ for a library of PaRC with distinct $L_e$ (Fig.~\ref{fig:Scaling}(a)). Additionally, in Fig.~\ref{fig:Scaling}(b) and (c) we plot $SF$ as a function of the reduced extensibility parameter $\tilde{L_e}=( \langle l_c \rangle - \sqrt{\langle R^2\rangle})/\sqrt{\langle R^2\rangle}= L_e -1$, to highlight the differences for stiffer PaRC, and as a function of $N_p$, respectively. 
Notwithstanding the different chemical structures and architectures of the PaRC investigated, a general trend of $SF$ as a function of $L_e$ emerges. Specifically, the $SF$ increase with $L_e$ can be approximated as $SF= 2 L_e^{\beta}$, where $\beta=2/3$ is a dimensionless parameter (blue line, Fig.~\ref{fig:Scaling}(a)). This function also captures the main features for the $SF$ as a function of $\tilde{L_e}$ and $N_p$ in Fig.~\ref{fig:Scaling}(b) and (c), respectively. The trend of $SF$ as a function of $N_p$  shows a plateau for $N_p<1$ and a transition towards $SF \propto N_p^{1/3}$ at $N_p \gtrsim 10$ (Fig.~\ref{fig:Scaling}(b)). 
We physically interpret the evolution of $SF$ as a function of $L_e$, $\tilde{L_e}$ and $N_p$ based on the flow-induced conformational changes occurring to the PaRC under the contrasting flow types.
Rigid PaRC ($L_e\sim1$ or $N_p\ll 1$) preserve their rod-like conformation independently of the flow strength and type. Thus, the effectiveness of extensional flows at inducing PaRC alignment ($SF\sim2$) can be exclusively associated with the absence of tumbling events in extensional flows. However, more extensible or flexible PaRC  ($L_e >1$, $N_p\gtrsim 1$) have the ability to adopt different conformations according to the flow strength and type (e.g., dumbbell, hairpin, etc)~\cite{smith1999single,kirchenbuechler2014direct}. 
In shear flows intricate conformations and dynamics can occur due to the vorticity component, minimizing the overall degree of backbone orientation. This is supported by single-polymer studies in shear flows, reporting chain extension fluctuation for DNA with $N_p > 1$, phenomena not observed in extensional flows~\cite{smith1999single,Schroeder2005}.
Additionally, shear flow has been found to stabilize hairpin conformation for actin filaments with $N_p\sim 1$ whilst an extended conformation has been reported in extensional flows~\cite{kirchenbuechler2014direct,Liu2020}. As such, for PaRC with $N_p\gtrsim 1$, the enhanced efficacy of extensional deformations at inducing fluid anisotropy, captured by $SF>2$, stems from a more extended conformation occurring in extensional flows compared to shear flows. Note that most experiments were performed above the overlap concentration in order to have sufficient birefringence \cite{supp}. We have tested experimentally that interparticle interactions in the semi-dilute regime do not affect the SF. This is also confirmed by theory for stiff PaCR\cite{supp}. For more flexible PaRC (e.g, Pf1) the underlying reason of the concentration independent SF needs theoretical underpinning.

Given the key role that PaRC play in many biological and industrial processes (e.g., mucus flow, inkjet printing), there is growing research devoted to studying their dynamics in shear and extensional flows (see e.g.,~\cite{dinic2022rheology,lindner2022morphological,Schroeder2018,calabrese2023naturally} and references therein). Here, we experimentally investigate the dynamics of multiple PaRC with different chemistry and architectures to provide a comprehensive understanding of their dynamics in distinct flow conditions. Our results show that the disparity between the critical deformation rates required for the onset of PaRC alignment in shear and extension is related to PaRC extensibility. We anticipate that our framework of analysis and derived scaling concepts will be tested by numerical simulations and prove beneficial for synthesising materials with anisotropic microstructures.

\bibliographystyle{ieeetr}







\end{document}